\documentclass{article}
\usepackage{spconf}
\usepackage{cite}
\usepackage{epstopdf}
\usepackage{stfloats}
\usepackage[cmex10]{amsmath}
\usepackage{amssymb}
\usepackage{graphicx}
\usepackage{multirow}
\usepackage{enumitem}
\usepackage{lipsum}
\usepackage{array}

\newcolumntype{C}[1]{>{\centering}m{#1}}
\usepackage[tight,footnotesize]{subfigure}
\usepackage{booktabs}
\usepackage{caption}
\date{}

\usepackage{enumitem}

\usepackage{url}

\usepackage{hyperref} 

\begin{document}
	
	\onecolumn
	
	\begin{description}[labelindent=0cm,leftmargin=3cm,rightmargin=3cm,style=multiline]
		
		\item[\textbf{Citation}]{M. Alfarraj and G. AlRegib, "Petrophysical property estimation from seismic data using recurrent neural networks," SEG Technical Program Expanded Abstracts 2018. Society of Exploration Geophysicists, 2018. 2141-2146.}
		
		
		\item[\textbf{Review}]{Date of presentation: 16 Oct. 2018}
		
		\item[\textbf{Data and Codes}]{\href{https://github.com/olivesgatech/Petrophysical-property-estimation-from-seismic-data-using-recurrent-neural-networks}{[\underline{GitHub Link}]}}

		\item[\textbf{Bib}] {@incollection\{alfarraj2018petrophysical,\\
  title={Petrophysical property estimation from seismic data using recurrent neural networks},\\
  author={Alfarraj, Motaz and AlRegib, Ghassan},\\
  booktitle={SEG Technical Program Expanded Abstracts 2018},\\
  pages={2141--2146},\\
  year={2018},\\
  publisher={Society of Exploration Geophysicists\}\\
}}

		
		\item[\textbf{Contact}]{\href{mailto:motaz@gatech.edu}{motaz@gatech.edu}  OR \href{mailto:alregib@gatech.edu}{alregib@gatech.edu}\\ \url{http://ghassanalregib.com/} \\ }
	\end{description}
	
	\thispagestyle{empty}
	\newpage
	\clearpage
	\setcounter{page}{1}
	
	\twocolumn
	
\title{Petrophysical property estimation from seismic data using recurrent neural networks}
\name{Motaz Alfarraj and Ghassan AlRegib}
\address{Center for Energy and Geo Processing (CeGP) \\ School of Electrical and Computer Engineering \\ Georgia Institute of Technology \\ \{motaz,alregib\}@gatech.edu}

\maketitle

\begin{abstract}
Reservoir characterization involves the estimation petrophysical properties from well-log data and seismic data. Estimating such properties is a challenging task due to the non-linearity and heterogeneity of the subsurface. Various attempts have been made to estimate petrophysical properties using machine learning techniques such as feed-forward neural networks and support vector regression (SVR). Recent advances in machine learning have shown promising results for recurrent neural networks (RNN) in modeling complex sequential data such as videos and speech signals. In this work, we propose an algorithm for property estimation from seismic data using recurrent neural networks. An applications of the proposed workflow to estimate density and p-wave impedance using seismic data shows promising results compared to feed-forward neural networks.  
\end{abstract}

\section{Introduction}
Reservoir characterization (RC) is the process of estimating petrophysical properties of the subsurface using information obtained from well-log, core, and seismic data. The goal of RC is to estimate petrophysical properties such as porosity, density and permeability at any location and depth in a reservoir. RC is a complex process due to the non-linearity and heterogeneity of the subsurface. There is no clear mapping from seismic data to well-logs, and even if such mapping exists it might not generalize well beyond the study area.

\begin{figure}[ht]
    \centering
    \includegraphics[width=\linewidth]{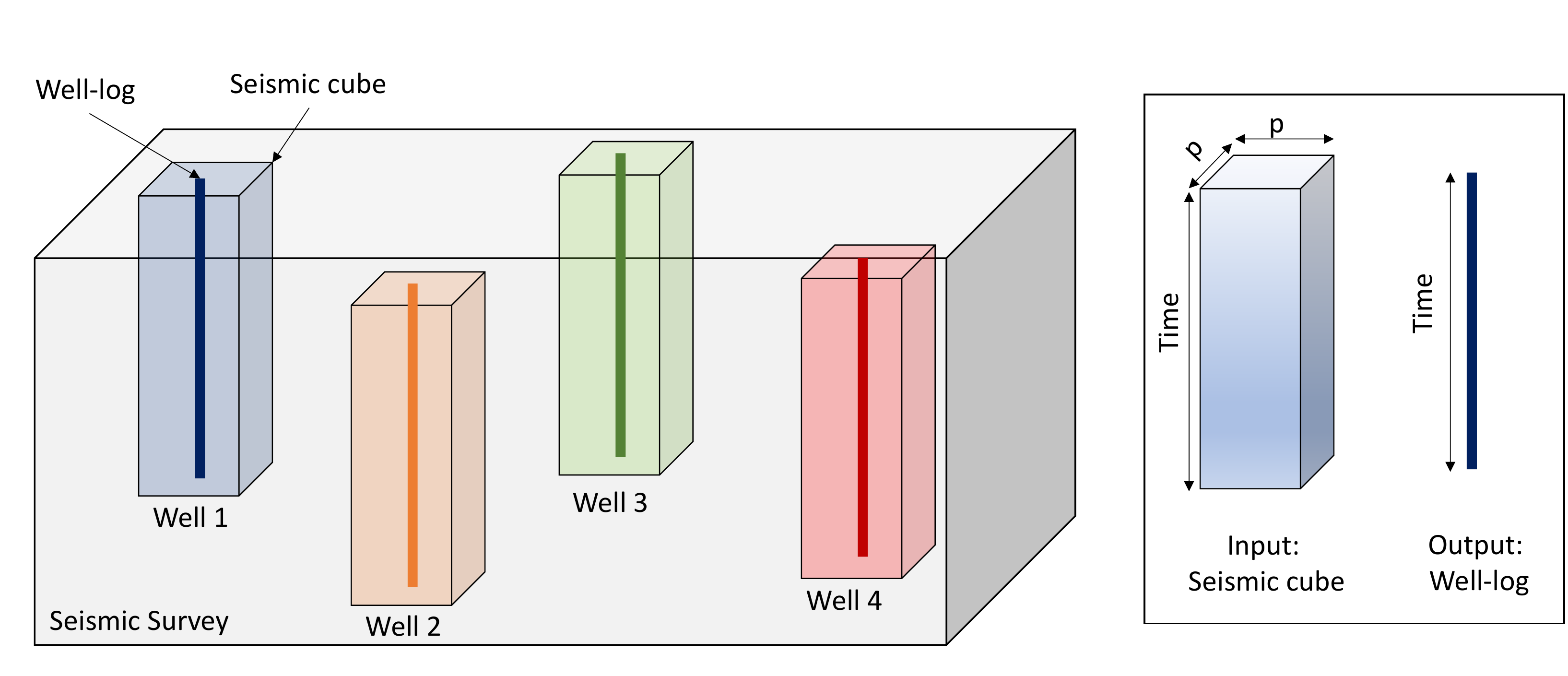}
    \caption{An illustration of the integration of well log data and seismic data in a survey area.}
    \label{fig:survey}
\end{figure}

Simply stated, the RC problem is finding a functional approximation from seismic data to well-log data so that log data can be generalized beyond well location to the entire reservoir area. From a machine learning perspective, the goal is to train an estimation model on the sparsely available well-logs and their corresponding seismic data (as illustrated in Figure \ref{fig:survey}) such that it can estimate one or several well-logs properties at a given location and depth/time using seismic data at the same location. Then, the model can be used to generate a property volume for the entire reservoir area. 

Although this problem might seem to be a perfect setup for regression algorithms such as support vector regression (SVR), decision trees, and feed-forward neural networks, there are many challenges that prevent such algorithms to find a proper mapping that can be generalized for an entire survey area. One of the challenges is the lack of data from a given survey area on which a model can be trained, as we are limited to the number of drilled wells in an area. For this reason, such regression algorithms need to have a limited number of parameters and a good regularization mechanism in order to prevent over-fitting and to be able to generalize beyond the training data. In addition, there are two common methods to model the problem so that regression algorithms can be used. The first method is to treat each data point in a well-log (in depth) as an independent sample and try to estimate its value from the corresponding seismic data sample(s). This method fails to capture the temporal dynamics of well-log data that is the dependency of a data point at a given depth on the data points before it and after it. An alternative approach is to estimate the entire well-log at once from the corresponding seismic data to incorporate the temporal dependency (in depth/time) of petrophysical properties. However, this approach severely limits the amount of data from which the algorithm can learn; because each well-log in this scheme is treated as a single training sample. With a limited amount of data samples, common machine learning algorithms will fail to generalize beyond the training data. Furthermore, seismic data are captured at lower resolution than that of well-log data which make this problem even more difficult. In order to remedy this issue, a data preprocessing step is required before attempting to train any machine learning algorithms \cite{chaki2018well}. 

Several attempts have been made using machine learning and statistical learning tools such as artificial neural networks, and support vector regression to solve the RC problem \cite{al2012support, chaki2015novel,gholami2017estimation,chaki2017diffusion}. The literature shows great promise for machine learning algorithms for property estimation. However, most regression algorithms treat data samples independently such that a prediction is made solely from the input data with no influence from the outputs from data points before or after the target point. Well-log data exhibit inter-log correlations, such that logs may follow certain intrinsic patterns due to consistency in lithology in a given study area. Furthermore, well-logs also exhibit inter-log (temporal) correlations, i.e. correlations between property samples for a given depth range. In this study, we propose the use of recurrent neural networks (RNNs) to capture the aforementioned correlations of wells logs in a given survey area by modeling well-log data as sequences (in depth/time). The proposed workflow is trained and validated using well-logs and their corresponding seismic data from the Netherlands offshore F3 block. 

\section{Feed-forward and Recurrent Networks}
Despite the success of feed-forward neural networks for various learning tasks, they have their limitations. Feed-forward neural networks have an underlying assumption that data points are independent and thus the internal state of the networks is cleared after a data sample is processed which would be fine, unless data is not independent which is the case for sequential data.

Recurrent neural networks are a class of artificial neural networks that can capture temporal dynamics of sequential data like time series, audio and video. Unlike feed-forward neural networks, RNNs have a hidden state that can be passed between sequence samples which serves as memory allowing them to capture very long temporal dependencies in sequential data. RNNs have often been utilized to solve many problems in language modeling and natural language processing (NLP)\cite{mikolov2010recurrent}, speech and audio processing \cite{graves2013speech}, and video processing \cite{ma2017ts}.

A single layer feed-forward neural network produces an output $\mathbf{y}_i$ which is a weighted sum of input features $\mathbf{x}_i$ followed by an activation function (a non-linearity) like the sigmoid or hyperbolic tangent functions, i.e. $\mathbf{y}_i = \sigma\left(\mathbf{W}\mathbf{x}_i+\mathbf{b}\right),$ where $\mathbf{x}_i$  and $\mathbf{y}_i$ are the input and output feature vectors of the $i^{th}$ sample, respectively, $\sigma(\cdot)$ is the activation function, $\mathbf{W}$ and $\mathbf{b}$ are the learnable weights matrix and bias vector, respectively. The same equation is applied for all data samples independently to produce outputs. 

\begin{figure}[h]
    \centering
    \includegraphics[width=0.85\linewidth]{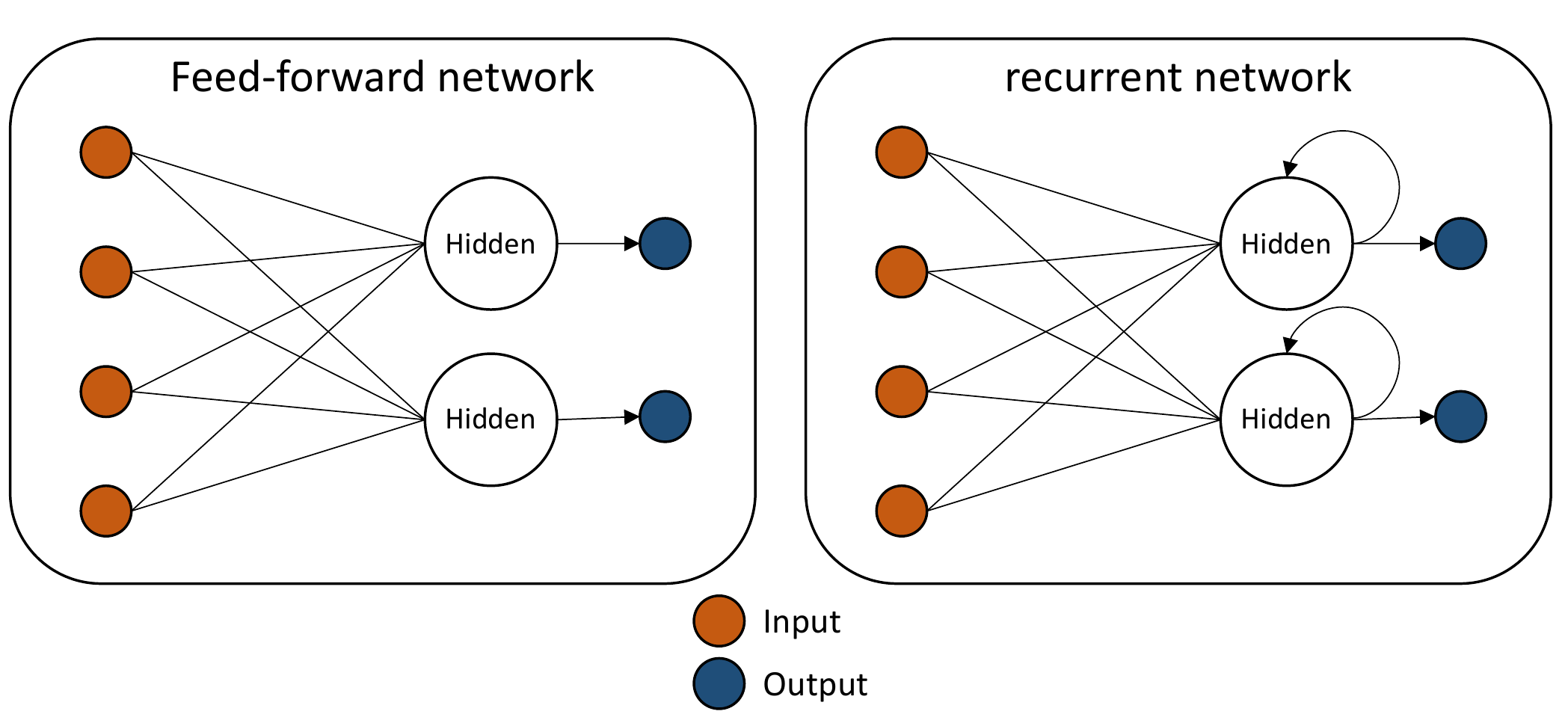}
    \caption{An illustration of feed-forward and recurrent networks.}
    \label{fig:RNN_vs_NN}
\end{figure}

In addition to the affine transformation and non-linearity, RNNs introduce a hidden state variable that is computed using the current input and the hidden state variable from the previous step, 
\begin{equation}
    \begin{aligned}
    \mathbf{h}_i^{(t)} &= \sigma\left(\mathbf{W}_{xh}\mathbf{x}_i^{(t)}+\mathbf{W}_{hh}\mathbf{h}_i^{(t-1)} + \mathbf{b}_h\right), \\
    \mathbf{y}_i^{(t)} &= \sigma\left(\mathbf{W}_{hy}\mathbf{h}_i^{(t)} + \mathbf{b}_y\right) 
    \end{aligned}
    \label{eqn:RNN}
\end{equation}

where $\mathbf{x}_i^{(t)}$, $\mathbf{y}_i^{(t)}$ and $\mathbf{h}_i^{(t)}$, are the input, output, and state vectors at time step $t$, respectively, $\mathbf{W}$'s and  $\mathbf{b}$'s are network weights, and bias vectors respectively. For time $t=0$, the hidden state variable is set to $\mathbf{h}^{(0)} = \mathbf{0}$. Figure \ref{fig:RNN_vs_NN} shows a side-by-side comparison between a feed-forward unit and a recurrent unit. 

When RNNs were first proposed in 1980s, they were hard to train because they introduced a dependency between data samples which made the gradients more difficult to compute. Additionally, they have more parameters to learn compared to feed-forward networks. The problem was solved using backpropagation through time (BPTT) algorithms \cite{werbos1990backpropagation}, which turns gradients into a long product of terms using the chain rule. Theoretically, RNNs are supposed to learn long-term dependencies from their hidden state variable. However, even with BPTT, RNNs failed to learn long-term dependencies mainly because the gradients tend to either vanish or explode for long sequences as they were backpropagated through time.

New RNN architectures with more sophisticated activation functions have been proposed to overcome the issue of vanishing gradients using gated units. Examples of such architectures are Long Short-Term Memory (LSTM) \cite{hochreiter1997lstm} and the recently proposed Gated Recurrent Units (GRU) \cite{cho2014properties}. Such architectures have been shown to capture long-term dependency and perform well for various tasks such as machine translation and speech recognition. In this paper, we utilize GRUs in our proposed model to enhance the estimation of petrophysical properties from seismic data. 

\subsection{Gated Recurrent Units}

\begin{figure}[h]
    \centering
    \includegraphics[width=0.9\linewidth]{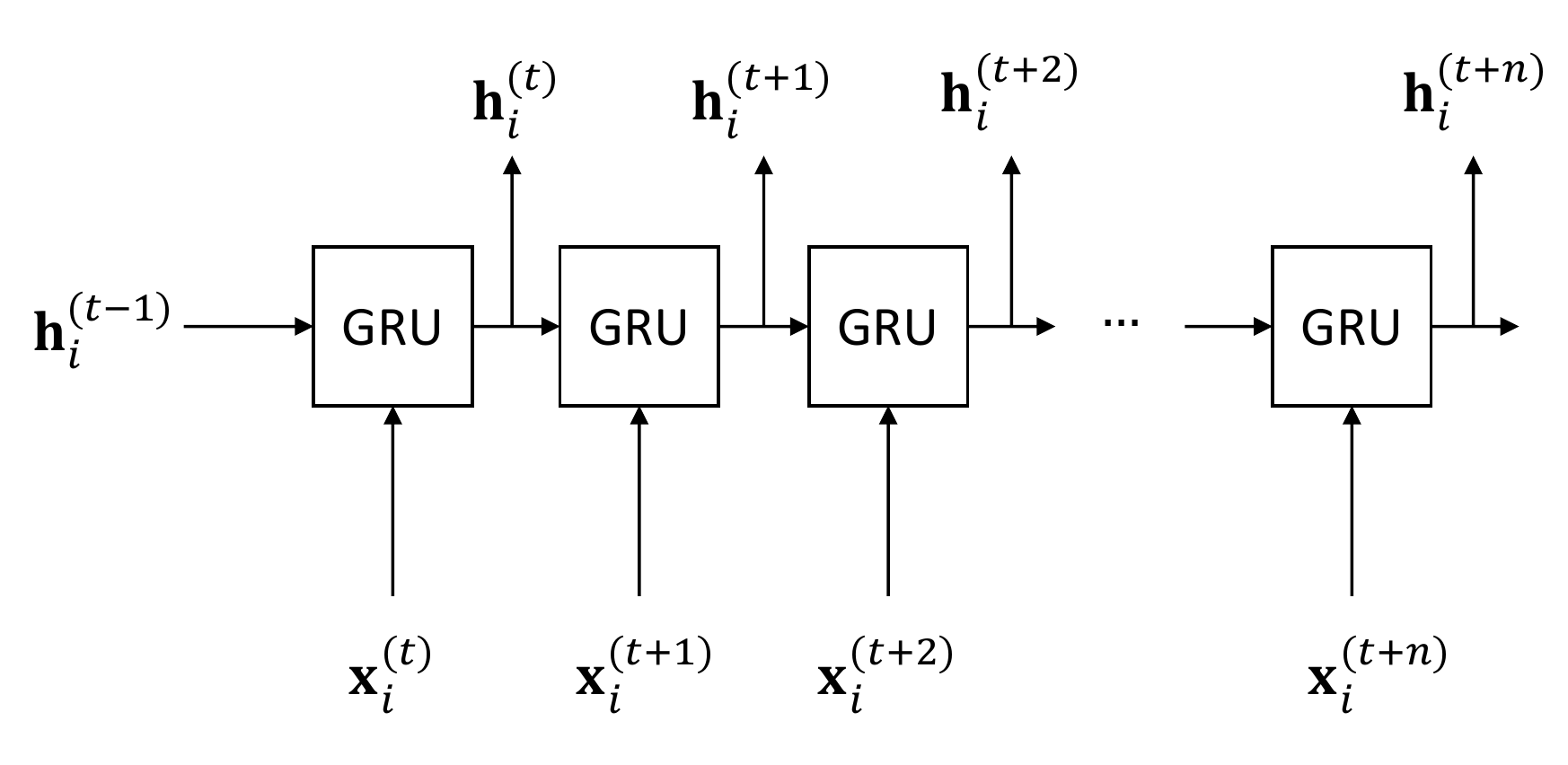}
    \caption{Gated Recurrent Unit (GRU) unfolded through time.}
    \label{fig:GRU}
\end{figure}

“GRUs supplement the simple RNN described above by incorporating a reset-gate and an update-gate variables which are internal states that are used to evaluate the long-term dependency and keep information from previous times only if they are needed. The forward step through a GRU is given by the following equations, 
\begin{equation}
    \begin{aligned}
        \mathbf{u}_i^{(t)} &= \text{sigmoid}\left(\mathbf{W}_{xu}\mathbf{x}_i^{(t)}+\mathbf{W}_{hu}\mathbf{h}_i^{(t-1)} + \mathbf{b}_z\right) \\
        \mathbf{r}_i^{(t)} &= \text{sigmoid}\left(\mathbf{W}_{xr}\mathbf{x}_i^{(t)} + \mathbf{W}_{hr}\mathbf{h}_i^{(t-1)} + \mathbf{b}_r\right) \\
        \hat{\mathbf{h}}_i^{(t)} &=  \tanh\left(\mathbf{W}_{x\hat{h}}\mathbf{x}_i^{(t)} + \mathbf{b}_{\hat{h}_1}+  r_i^{(t)}\circ\left(\mathbf{W}_{h\hat{h}}\mathbf{h}_i^{(t-1)} +  \mathbf{b}_{\hat{h}_2}\right)\right)\\
        \mathbf{h}_i^{(t)} &= (1-\mathbf{u}^{(t)})\circ \mathbf{h}_i^{(t-1)} + \mathbf{u}^{(t)}\circ \hat{\mathbf{h}}^{(t)}
    \end{aligned}
    \label{eqn:GRU}
\end{equation}

where $\mathbf{z}_i^{(t)}$ and $\mathbf{r}_i^{(t)}$ are the update-gate, and reset-gate vectors, respectively, $\hat{\mathbf{y}}_i^{(t)}$ is the candidate output, $\mathbf{W}$'s and $\mathbf{b}$'s are the learnable parameters, and $\circ$ is the element-wise product. Note that in addition to the output state, the GRU introduces two new state variables, update-gate $\mathbf{u}$ and reset-gate $\mathbf{r}$, which control the flow of information from one time step to another, and thus they are able to capture long-term dependency. Figure \ref{fig:GRU} shows an example of a GRU network unfolded through time. Note that all GRU's in an unfolded network share the same parameters.

\section{Method}

\subsection{Data Preprocessing}
Well-logs are acquired at a much higher vertical resolution than seismic attributes which requires a preprocessing step in order to successfully train an estimation model and guarantee its convergence. One approach to preprocessing the data is to regularize the logs by smoothing such that both the logs and seismic attributes have comparable information content \cite{chaki2018well}. This is done by filtering log data with a low-pass filter to match frequency content of seismic data. This step reduces the variation of log data in a small time window so that the model can capture the overall trend of logs rather than the small high frequency variations. Furthermore, the data samples are normalized such that each log trace has a zero mean and a unit standard deviation which is a common step before training a machine learning model. 

\subsection{Proposed Model}
\begin{figure}[h]
    \centering
    \includegraphics[width=\linewidth]{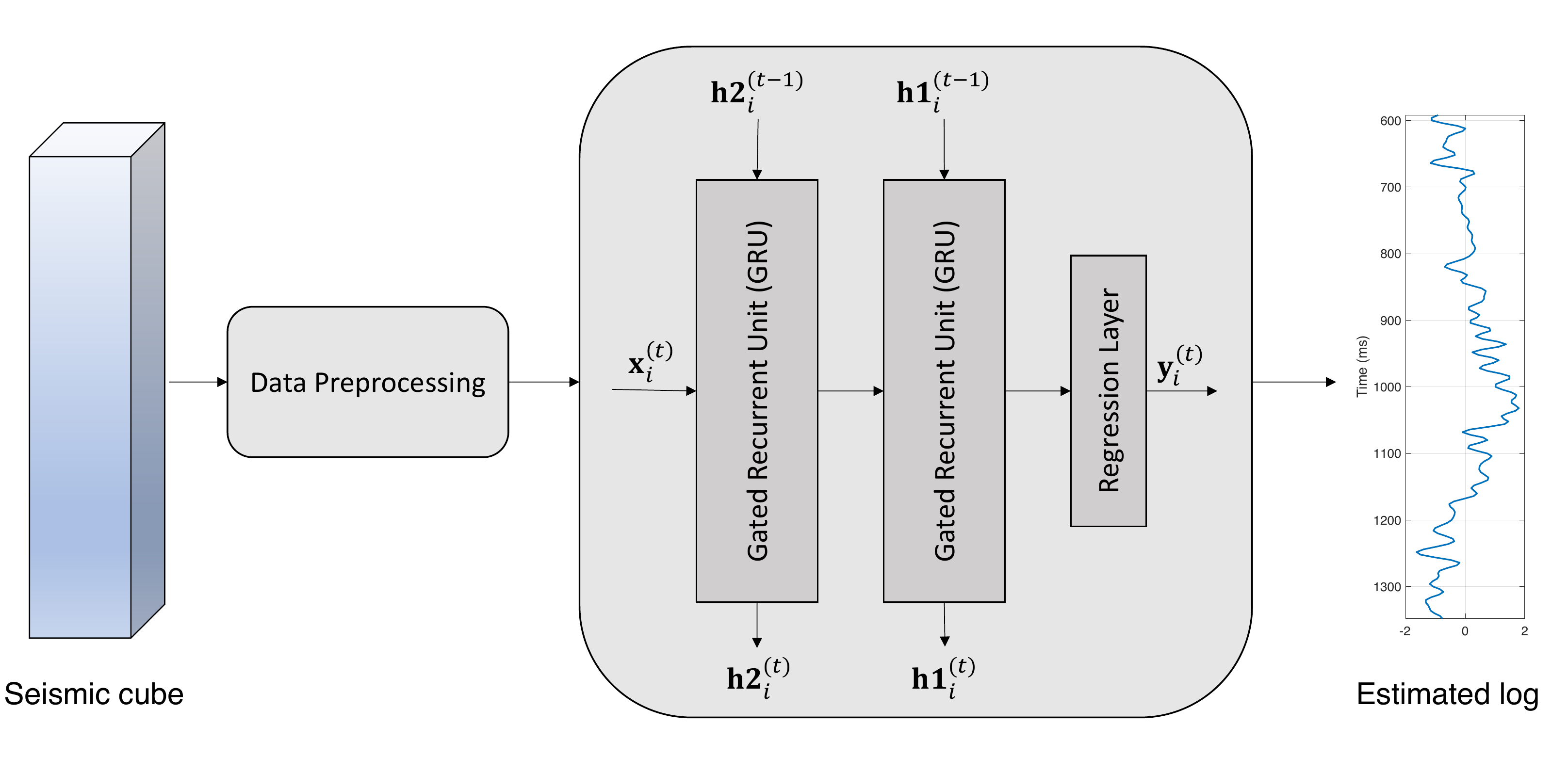}
    \caption{The proposed workflow with 2 layer GRU and a regression layer.}
    \label{fig:proposed_workflow}
\end{figure}

In order to capture the inter- and intra- log correlations as well as to establish a functional approximation from seismic to log data, we propose a simple 2-layer recurrent neural network, namely a GRU, followed by a linear regression layer. As we have discussed above, outputs of the GRU are a function of an affine transform of the inputs plus bias, which can be seen as feed-forward network by itself. In addition, it utilizes the update-gate and reset-gate variables to improve the network's outputs at a given time step based on the networks previous states. The proposed workflow is shown in Figure \ref{fig:proposed_workflow}. 

For a given well log, a seismic cube is extracted around the well location to be used as an input to train the model. The seismic cube is of size $p\times p\times T$ where $p$ is the number of seismic traces in the inline and crossline directions, and $T$ is the number of samples in a trace. Let $\mathbf{x}_i \in \mathbb{R}^{p\times p\times T}$ be the seismic cube at location $i$, and $\mathbf{y}_i$ be the log trace at the same location. The model processes the data sequentially (in time) such that it inputs the seismic slice at time $t$, $\mathbf{x}_i^{(t)}\in \mathbb{R}^{p\times p}$, and the state variables of both GRUs at time $t-1$, $\mathbf{\Tilde{h}1}_i^{(t-1)}$ and $\mathbf{\Tilde{h}2}_i^{(t-1)}$, in order to compute the output state variables at time $t$. The regression layer then takes $\Tilde{\mathbf{h2}}_i^{(t)}$ and computes the estimated property at time $t$, $\mathbf{\Tilde{y}}_i^{(t)}$. If the sample to be predicted is the first sample in the log ($t=0$), state variables are set to zero. The process is then repeated to estimate the entire property trace. During the training of the model, when all the $N$ logs in the training dataset have been estimated as $\tilde{\mathbf{y}}_i, \forall i=1,\dots,N$, they are compared to the measured log $\mathbf{y}_i,\forall i=1,\dots,N$ using Mean Squared Error (MSE) loss function. The error is then used to compute the gradients and to correct the model's parameters using BPTT. 

After proper training, the model's performance is assessed on the validation dataset by computing the Pearson correlation coefficient between the estimated logs and the measured logs. The Pearson correlation coefficient is computed as, 
%

\section{Experimental Evaluation}

The dataset contains 4 wells, \texttt{F02–1}, \texttt{F03–2}, \texttt{F03–4}, and \texttt{F06–1} from the Netherlands offshore F3 block. For each of the wells, we extracted a seismic cube of $7\times 7$ traces centered at the well ($p=7$ as in Figure \ref{fig:survey}). The proposed workflow is then trained using seismic cubes as inputs and a single property log from the well-log data. In our experiments, we trained two identical networks, one to estimate density and the other to estimate p-wave impedance, both of the networks are similar to the one shown in Figure \ref{fig:proposed_workflow}.

Due to the small size of the dataset, training regularization is needed to ensure that the model does not over-fit to the training data. One such technique is early stopping in which the training is stopped after a small number of epochs. More training epochs will definitely improve the performance of the model on the training dataset, but the model will fail to generalize. In addition, we used data augmentation by using multiple rotations of the seismic cubes along the time axis to increase the number of the training samples.

The model in Figure \ref{fig:proposed_workflow} with a 2 layer, 32-feature hidden state variable GRU was tested on the dataset described above. In addition, the same dataset was used to train a 2-layer, 32-neuron feed-forward neural network. The performance of the models is then assessed using 4-fold validation, where three of the wells are used for training and the remaining well is used for testing. The process is repeated 4 times, and the results are averaged for all experiments. The results are summarized in Table \ref{tab:correlation_results}. The results show that even with a small dataset, the recurrent neural network can estimate log data from seismic data with much higher correlation than the feed-forward network. Note that the feed-forward network was not able to train properly on such a small dataset.

\begin{table}[h]
    \centering
    \resizebox{\linewidth}{!}{
    \begin{tabular}{c|c|c|c|c}
         & \multicolumn{2}{c}{Feed-forward} & \multicolumn{2}{c}{Recurrent}\\
        Property& Training & Validation &Training & Validation \\
        \hline
        P impedance & 0.48 & 0.37 & 0.96 & 0.72\\ 
        Density &  0.42 & 0.31& 0.97 & 0.70 \\
        \hline
    \end{tabular}}
    \caption{Correlation coefficient between estimated and measured properties.}
    \label{tab:correlation_results}
\end{table}

\begin{figure}[h]
  \centering
    \includegraphics[width=0.9\linewidth]{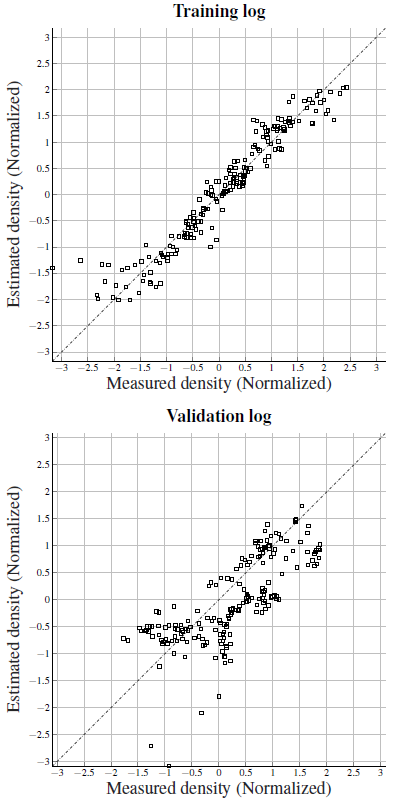}
    \caption{Scatter plots of measured density and estimated density from the training and validation datasets.}
    \label{fig:density_corr}
\end{figure}

\begin{figure}[ht]
    \centering
    \includegraphics[width=0.9\linewidth]{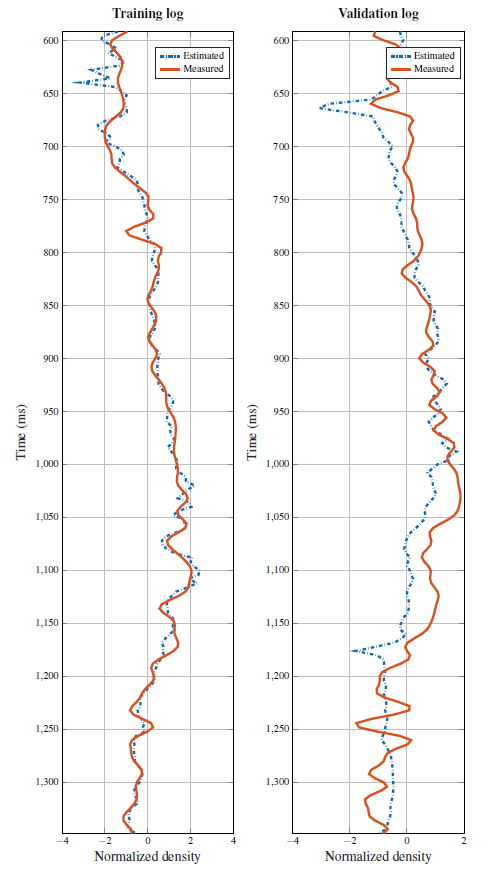}
    \caption{An example of measured density and estimated density logs from the training and validation datasets.}
    \label{fig:density_logs}
\end{figure}

Figure \ref{fig:density_corr} shows a scatter plot of the measured density and the estimated density using the proposed workflow for training and validation datasets. We can see that the estimated density varies almost linearly with respect to the measured density. Figure \ref{fig:density_logs} shows examples of estimated density logs using the proposed workflow. 
 
t is worth noting that a problem as difficult as property estimation might need a more complex and deeper learning model; however, the number of model parameters increase with complexity and thus much more data is required to train such models properly. The goal of this experiment was to show the power of recurrent neural networks for property estimation by utilizing their temporal dependencies, compared to the feed-forward neural networks which treat data samples independently. 

\section{Conclusions}
In this paper, we proposed a machine learning algorithm for well-log property estimation from seismic data using recurrent neural networks. The proposed workflow was validated using 4-fold validation for density and p-wave impedance estimation from seismic data. Although the training was carried out on a small dataset, the validation results indicate a great potential of recurrent neural networks for reservoir characterization. With a larger dataset for training, the model could be used to generate property volumes for a survey area from seismic data. 
\section{Acknowledgements}
This work is supported by the Center for Energy and Geo Processing (CeGP) at Georgia Institute of Technology and King Fahd University of Petroleum and Minerals (KFUPM).

\end{document}